\documentclass[10pt,twocolumn,letterpaper]{article}

\newcommand{\datasetname}{\textit{CONVEX}}

\usepackage{cvpr}             

\definecolor{cvprblue}{rgb}{0.21,0.49,0.74}
\usepackage[pagebackref,breaklinks,colorlinks,allcolors=cvprblue]{hyperref}

\usepackage[accsupp]{axessibility} 

\title{The Synthetic Media Shift: Tracking the Rise, Virality, and Detectability of AI-Generated Multimodal Misinformation
\thanks{Accepted at the 3rd Workshop on New Trends in AI-Generated Media and Security (AIMS), held in conjunction with CVPR 2026.}
}

\author{Zacharias Chrysidis, Stefanos-Iordanis Papadopoulos, Symeon Papadopoulos\\
Centre for Research and Technology Hellas, Greece\\
{\tt\small \{zchrysid, stefpapad, papadop\}@iti.gr}
}

\begin{document}
\maketitle

\begin{abstract}

As generative AI advances, the distinction between authentic and synthetic media is increasingly blurred, challenging the integrity of online information. In this study, we present \datasetname, a large-scale dataset of multimodal misinformation involving miscaptioned, edited, and AI-generated visual content, comprising over 150K multimodal posts with associated notes and engagement metrics from X's Community Notes. We analyze how multimodal misinformation evolves in terms of virality, engagement, and consensus dynamics, with a focus on synthetic media. Our results show that while AI-generated content achieves disproportionate virality, its spread is driven primarily by passive engagement rather than active discourse. Despite slower initial reporting, AI-generated content reaches community consensus more quickly once flagged. Moreover, our evaluation of specialized detectors and vision-language models reveals a consistent decline in performance over time in distinguishing synthetic from authentic images as generative models evolve. These findings highlight the need for continuous monitoring and adaptive strategies in the rapidly evolving digital information environment.

\end{abstract}
\section{Introduction}
\label{sec:introduction}

Amid the rapid evolution of media and communication technologies, the scale and velocity at which misinformation spreads have made it a major societal concern with potential negative impacts on
democratic processes \cite{bennett2018disinformation},
vulnerable groups \cite{gamir2021multimodal}, 
and public health \cite{do2022infodemics}, among other domains.
While early research largely focused on textual claims, misleading information increasingly incorporates multimodal content \cite{akhtar2023multimodal}, including images and videos, 
which tend to be perceived as more persuasive when used to support misleading claims or narratives \cite{weikmann2023visual}.
Generative AI amplifies these concerns by enabling the creation of realistic synthetic images, videos, and text at scale, raising questions about how AI-generated media may transform the production and spread of misinformation online \cite{dufour2024ammeba,augenstein2024factuality}.

Addressing misinformation at scale remains challenging for social media platforms.
Mitigation approaches rely on professional fact-checkers or automated detection systems \cite{augenstein2025epistemology}.
While expert verification can provide high-quality assessments, it struggles to scale to the massive volume of online content \cite{wirtschafter2023future}.
In turn, automated methods remain constrained by biases in training data, limited generalization and trustworthiness, and the rapidly evolving strategies used to produce misleading information \cite{deng2025survey, guo2022survey}.
Consequently, platforms are increasingly exploring community-based moderation, where users collaboratively contribute context and evaluate potentially misleading content \cite{borenstein2025cancnreplace}.

\begin{figure}[t] % 
    \centering
    \includegraphics[width=\columnwidth]{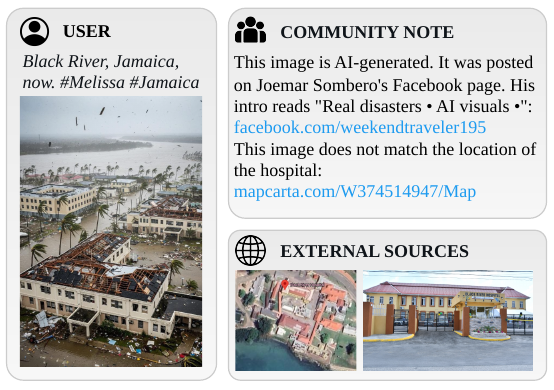} 
    \caption{Example of an AI-generated image post on X, with the Community Note and external sources used to verify it.}
    \label{fig:post_note_example}
\end{figure}

One prominent example is X's Community Notes, a crowdsourced fact-checking system introduced in 2021 that allows contributors to write contextual notes on posts they consider misleading, while other users rate their helpfulness \cite{wojcik2022birdwatchofficial}.
Figure~\ref{fig:post_note_example}
illustrates a community note, in which an AI-generated image is challenged through comparison with external satellite imagery, revealing inconsistencies between the depicted building and the stated location.
Since its introduction, Community Notes has 
expanded globally and 
produced millions of notes, providing a large-scale resource of community-annotated content \cite{mohammadi2025frombirdwatch}.
Prior research has utilized this data to analyze consensus dynamics \cite{wirtschafter2023future}, the timeliness of community responses \cite{razuvayevskaya2025timeliness}, and the longitudinal impact of the system on user engagement \cite{chuai2024rollout}.
However, the evolving landscape of multimodal misinformation and AI-generated media remains unexplored.

In this work, we leverage Community Notes to construct a large-scale dataset of multimodal misinformation -- categorized into miscaptioned, edited, and AI-generated images and videos -- and analyze their prevalence, engagement dynamics, and community-driven oversight. 
Specifically, we introduce \datasetname, the `Community Notes for Visual Misinformation on X' dataset,
a collection of over 150,000 note-post pairs with crowdsourced annotations, associated media, and engagement statistics. 
Notably, our data collection and annotation pipeline is designed to integrate future releases of Community Notes for continuous monitoring. 

We conduct a longitudinal data analysis of how different misinformation categories evolve in terms of virality, engagement, and consensus dynamics. 
Our analysis indicates that AI-generated content volume correlates with the evolution and availability of generative models. 
We find that generated media achieves disproportionate virality primarily through passive engagement (e.g., favorites), contrasting the more discursive patterns (e.g., replies) of miscaptioned content. 
Furthermore, while AI-generated media is initially slower to be reported, it currently exhibits significantly higher community consensus once identified.

Finally, motivated by the use of AI tools by Community Notes contributors, we construct a real-world benchmark of authentic versus AI-generated images.
Our evaluation of specialized Synthetic Image Detectors (SIDs) and Vision-Language Models (VLMs) reveals consistent and significant decline in detection efficacy over time as generative models continue to evolve.
We make the codebase, dataset, and appendix publicly available for reproducibility\footnote{\url{https://github.com/zachos99/convex-dataset}}.

\begin{figure*}[t!]
    \centering
    \begin{subfigure}{\columnwidth}
        \centering
        \includegraphics[width=\linewidth]{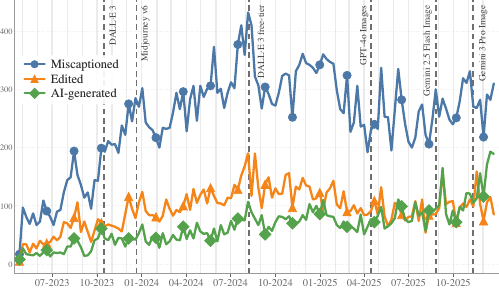}
        \caption{Image Set}
        \label{fig:volume_image}
    \end{subfigure}
    \hfill
    \begin{subfigure}{\columnwidth}
        \centering
        \includegraphics[width=\linewidth]{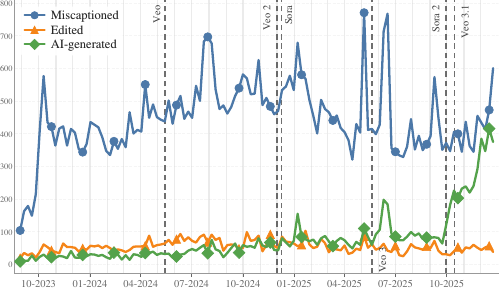}
        \caption{Video Set}
        \label{fig:volume_video}
    \end{subfigure}
    \caption{Monthly Community Notes volume for both modalities and releases of popular generative AI tools.}
    \label{fig:volume_combined}
\end{figure*}
\section{Related Work}
\label{sec:related_work}

\subsection{Multimodal and AI-Generated Misinformation}

Research on misinformation has increasingly recognized the importance of multimodal content, including images and videos that shape how users perceive and engage with claims online \cite{jin2017multimodal, akhtar2023multimodal}.
Recent studies show that manipulated or misleading visuals can be especially persuasive when presented as evidence of real-world events \cite{hameleers2020picture, weikmann2023visual}.  
Moreover, multimodal misinformation has become a substantial and evolving phenomenon, with empirical analyses indicating that contextual manipulations remain the most common form of visual misinformation even as AI-generated content has risen rapidly in recent years \cite{dufour2024ammeba,tonglet2025cove}. 
At the same time, the emergence of generative AI systems has increased concerns about the scalable production of realistic synthetic media and its implications for fact-checking and online information integrity  \cite{augenstein2024factuality,mirsky2021deepfakes}.

However, prior work primarily relies on synthetic data \cite{luo2021newsclippings, papadopoulos2023synthetic, muller2020multimodal} or fact-checking sites \cite{zlatkova2019fact, yao2023mocheg, papadopoulos2024verite}, which lack insight into how multimodal misinformation circulates `in the wild’. 
Moreover, existing social media datasets \cite{jin2016novel, boididou2015verifying, boididou2018verifying} are limited to a small number of events. 
This lack of scale and diversity limits their ability to generalize to the rapidly evolving misinformation landscape.

\subsection{Community-based Fact-Checking}

Community-based fact-checking systems leverage collective intelligence to complement expert content moderation \cite{augenstein2025epistemology,martel2024crowds}. 
A leading example is X's Community Notes, which has provided a valuable resource for studying real-world misinformation dynamics.
Early studies describe the transition to Community Notes and the system’s evolving design and infrastructure \cite{mohammadi2025frombirdwatch}. During the pilot period, internal tests report that users exposed to notes were 25–34\% less likely to like or repost misleading tweets \cite{wojcik2022birdwatchofficial}. 
However, an analysis after the platform-wide rollout find no significant reduction in engagement, likely because notes appear too late in the post’s lifecycle \cite{chuai2024rollout}.
Other research highlights limits to scalability and consensus formation: note production is highly concentrated (the top 10\% of contributors produce about 58\% of notes), while only around 11.5\% of proposed notes ultimately reach publication consensus \cite{razuvayevskaya2025timeliness}.

Another line of work examine political asymmetries in note language and behavior \cite{kuuse2025politicalbias}, the relationship between community-based moderation and professional fact-checking \cite{borenstein2025cancnreplace}, and patterns of source credibility and bias in the outlets cited within notes \cite{kangur2026whochecks}.
Recently, several studies have explored how LLMs could assist the production, ranking, or summarization of Community Notes \cite{li2025scalingLLMnotes,de2025supernotes}.
However, most research on Community Notes focuses on textual misinformation, while the evolution of multimodal misinformation over time remains largely unexplored.

\subsection{Detection of AI-Generated Media}

Detecting AI-generated images has become an active research area as generative models rapidly improve in realism.
Recent surveys describe a broad landscape of detection methods, spanning artifact-based, frequency-domain, representation-based, and multimodal reasoning approaches \cite{lin2024detecting, mahara2026aidetectionsurvey}. 
These works also highlight the difficulty of achieving robust detection across diverse generative models and real-world distribution shifts
\cite{karageogiou2024evolution}.
Recent specialized detectors, such as SPAI \cite{karageorgiou2025spai}, RINE \cite{koutlis2024rine} and B-Free \cite{guillaro2025bfree}, aim to improve robustness through spectral learning, intermediate visual representations, and bias-suppressing training. 
In parallel, VLMs have emerged as general-purpose systems for visual understanding and reasoning \cite{yin2024visualreasoning}, 
with recent work exploring their potential as reasoning-based tools for identifying synthetic media \cite{jia2024gptdeepfakes}.

Despite these advancements, previous studies largely rely on curated synthetic-image benchmarks rather than `in the wild' images. Our work addresses this gap by evaluating detection models on images drawn from Community-annotated posts on X.

\section{Dataset Construction}
\label{sec:dataset}

%-------------------------------------------------------------------------

\begin{figure*}[t]
    \centering
    \begin{subfigure}{\columnwidth}
        \centering
        \includegraphics[width=\linewidth]{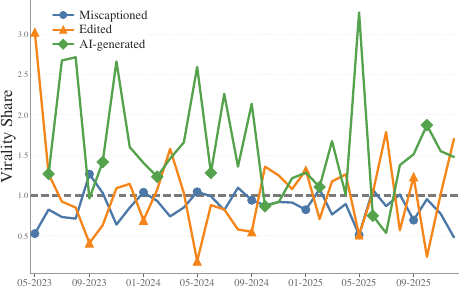}
        \caption{Image Set}
        \label{fig:p99_virality_image}
    \end{subfigure}
    \hfill
    \begin{subfigure}{\columnwidth}
        \centering
        \includegraphics[width=\linewidth]{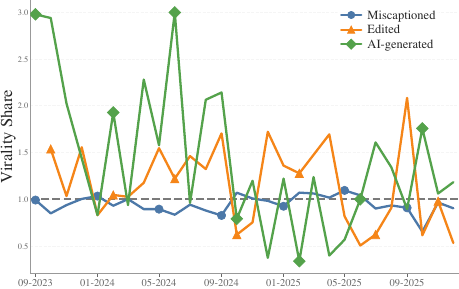}
        \caption{Video Set}
        \label{fig:p99_virality_video}
    \end{subfigure}
    
    \caption{Virality Share across both modalities.}
    \label{fig:p99_virality_main}
\end{figure*}

\subsection{Data Collection}

To construct \datasetname\ for longitudinal analysis of multimodal misinformation, we use the publicly available Community Notes corpus\footnote{\url{https://x.com/i/communitynotes/download-data}}. 
We collect all notes and metadata from January 2021 to January 2026 and retain entries labeled as `\emph{misinformed or potentially misleading}', resulting in 1,806,168 notes.

To isolate multimodal misinformation, we identify notes that reference images or videos. 
We employ keyword-based filtering (e.g., `photo', `graphic' for images; `clip', `footage' for videos; see 
Appendix 10 for full list) to identify multimodal entries. 
We then retrieve the associated X posts using the \texttt{twikit} library\footnote{\url{https://twikit.readthedocs.io}} and collect the corresponding media files, author metadata, and engagement metrics (favorites, retweets, views). 
Inaccessible posts (deleted or suspended) are excluded. 
The final corpus comprises 66,135 image-related and 86,131 video-related note–post pairs.

%-------------------------------------------------------------------------

\subsection{Data Annotation}

We classify note-post pairs into three misinformation categories: \textbf{Miscaptioned} (authentic media presented in misleading contexts), 
\textbf{Edited} (digitally altered media), 
and \textbf{AI-generated} (synthetic media).
To annotate the dataset at scale, we adopt a hybrid weakly supervised approach:
\begin{itemize}
    \item \textbf{Keyword-based}: For each category, we define a keyword list (e.g., `out-of-context', `reused photo' for miscaptioned; `photoshopped', `digitally altered' for edited; `AI-generated', `synthetic image' for AI-generated) 
    and apply them to the Community Note. % text. 
    \item \textbf{VLM-based}: 
    We use Gemma 3 in a zero-shot setting to classify each entry using the post text, associated media, and Community Note text. 
    \item \textbf{Classification}: When keyword-based and VLM-based labels disagree, we rerun the VLM with the keyword-derived label provided as additional context. 
    Final labels are assigned via majority voting over keyword-based labels and the two VLM predictions.
\end{itemize}
See Appendix 11 for the full methodology and prompts.

In the image subset (66K entries), 60.2\% are classified as miscaptioned, 23.3\% as edited and 16.3\% as AI-generated. 
In the video subset (86K entries), 75.6\% are classified as miscaptioned, 12.8\% as AI-generated, and 9.3\% as edited.
Few ambiguous cases were excluded from the analysis.
Although this weakly supervised procedure may introduce labeling noise, it enables consistent annotation 
at scale which is necessary for longitudinal analysis.

To assess label quality, we manually examine a stratified random sample of 600 note-post pairs (100 per category-modality combination), with weights reflecting the 2023–2025 distribution. Agreement with manual labels was 91\%, 95\%, and 92\% for AI-generated, edited, and miscaptioned images, versus 87\%, 88\%, and 91\% for videos.

\subsection{Continuous Monitoring}

To ensure continuous tracking and long-term analysis, our data collection and annotation pipeline is designed to operate on successive Community Notes releases. 
This enables monitoring shifts in the multimodal misinformation landscape over time rather than relying on static snapshots.

\section{Evolution of Multimodal Misinformation}
\label{sec:volume}

We examine monthly Community Notes volume for image- and video-related 
entries beginning in May 2023 and September 2023, respectively, 
following the platform's official media annotation rollout. 
Figure~\ref{fig:volume_combined} shows the
evolution of notes categorized as Miscaptioned, Edited and AI-generated across both modalities.
Miscaptioned content remains the dominant category for both images and videos, reflecting the low barrier to entry for repurposing authentic visual media. 
While edited content volume is relatively stable, AI-generated visual content displays a steady upward trajectory, with accelerated growth in recent periods -- most notably within the video subset.

These surges often align with major generative model releases. 
In the image subset, volume increases correspond with the public release of DALL-E 3 (August 2024) and the integration of GPT-4o image generation into free tiers (April 2025), or more recently, Gemini 3 Pro Image (``Nano Banana Pro''). 
Similarly, video notes spiked following the public releases of Sora 2 and Veo 3.1.
These patterns suggest that expanded access to high-fidelity generative tools correlates with the volume of AI-generated visual content.
\begin{figure*}[t]
     \centering
     \begin{subfigure}[b]{\columnwidth}
         \centering
         \includegraphics[width=\linewidth]{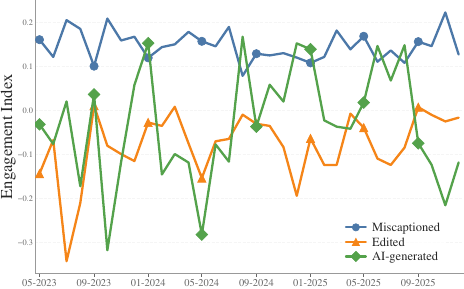}
         \caption{Image Set}
         \label{fig:engagement_image}
     \end{subfigure}
     \hfill
     \begin{subfigure}[b]{\columnwidth}
         \centering
         \includegraphics[width=\linewidth]{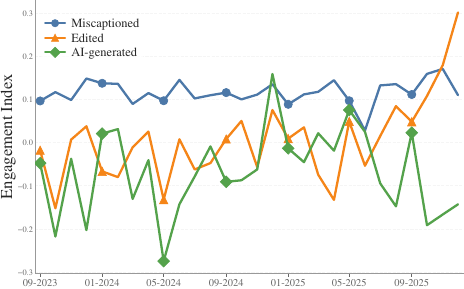}
         \caption{Video Set}
         \label{fig:engagement_video}
     \end{subfigure}
     \caption{Monthly Engagement Index for both modalities.}
     \label{fig:engagement_main}
\end{figure*}

\section{Attention Dynamics}
\label{sec:engagement}

%-------------------------------------------------------------------------
\subsection{Virality Share}

To assess how different types of misinformation capture public attention, we analyze engagement metrics, including retweets ($rt$), favorites ($f$), and replies ($rp$). 
We define an aggregate interaction score:
\begin{equation}
A = \text{rt} + \text{f} + \text{rp}
\end{equation} 
To account for the heavy-tailed nature of social media interactions, we define a post as \emph{viral} if its score $A$ exceeds the 99th percentile of the monthly distribution for its modality.

To compare the viral potential of each misinformation category while accounting for its baseline prevalence, we define a Virality Share metric:
\begin{equation}
V(c,m) = \frac{P(c \mid \text{viral}, m)}{P(c \mid m)}
\end{equation}
where $c$ denotes the misinformation category and $m$ the month.
This metric measures whether a given category is over- or under-represented among viral posts relative to its overall frequency. 
$V \approx 1$
indicate proportional representation, 
$V > 1$
indicate over-representation, and 
$V < 1$
indicate under-representation.

AI-generated content exhibits the highest average $V$ in both modalities, reaching 1.56 in the image subset and 1.25 in the video subset. 
Edited content remains close to proportional representation (1.02 image, 1.13 video), while miscaptioned posts are slightly under-represented among viral posts (0.91 image, 0.97 video). 
These results indicate that AI-generated content is disproportionately represented among highly engaging posts.

As shown in Figure \ref{fig:p99_virality_main}, AI-generated content is consistently over-represented among viral posts across most months in both modalities, with pronounced spikes in mid-2024 and mid-2025. 
While month-to-month volatility is expected given the small number of posts in the top 1\%, the overall pattern indicates persistent over-representation rather than isolated outliers. 
In contrast, miscaptioned content remains close to proportional representation ($V \approx 1$), and edited content fluctuates around the baseline.

\subsection{Engagement Dynamics}
Beyond virality, we examine user engagement; the ratio of active (retweets and replies) versus passive (favorites) engagement.
To smooth variance in the highly skewed engagement metrics on X, we apply a log transformation to each signal $s$:
\begin{equation}
l = \log(1 + s)
\end{equation}
We then standardize these values within each month using a z-score:
\begin{equation}
z = \frac{l - \mu_m}{\sigma_m}
\end{equation}
where $\mu_m$ and $\sigma_m$ denote the monthly mean and standard deviation. 
This normalization places all engagement signals on a comparable, zero-centered scale. 
Using the standardized signals, we define an Engagement index ($E$) as:
\begin{equation}
E = z_\text{rt} + z_\text{rp} - 2\,z_\text{f}
\label{eq:debate_index}
\end{equation}
The factor of two assigns equal total weight to active and passive 
interactions. Higher values indicate posts that trigger discursive participation--where users are more likely to retweet a post or reply than to simply ``like'', while lower values reflect passive approval. 

Figure \ref{fig:engagement_main} illustrates monthly median $E$ values across misinformation categories.
In both modalities miscaptioned content consistently shows higher median $E$ values than edited and AI-generated content. 
In contrast, AI-generated posts exhibit lower values, reflecting attention driven primarily by passive engagement rather than active discussion. 
These results suggest that miscaptioned content frequently sparks public discussion or controversy, while AI-generated content tends to propagate through passive engagement.
\section{Consensus Dynamics}
\label{sec:consensus}

\begin{table*}[t]
\centering
\setlength{\tabcolsep}{3pt}
\renewcommand{\arraystretch}{0.95}
\caption{Consensus metrics across misinformation categories for image and video subsets.}
\label{tab:consensus}
\begin{tabular}{l|ccccc|ccccc}
\hline
 & \multicolumn{5}{c|}{Image} & \multicolumn{5}{c}{Video} \\
Type  
& \shortstack{Notes / \\ Tweet}
& \shortstack{Helpful \\ (\%)}
& \shortstack{Consensus \\ Probability}
& \shortstack{First Note \\ (h)}
& \shortstack{Notes to \\ Consensus}
& \shortstack{Notes / \\ Tweet}
& \shortstack{Helpful \\ (\%)}
& \shortstack{Consensus \\ Probability}
& \shortstack{First Note \\ (h)}
& \shortstack{Notes to \\ Consensus}
\\
\hline
Miscaptioned 
& 1.89 & 66.8 & 0.284 & 9.41 & 1.47
& 1.85 & 71.1 & 0.295 & 9.06 & 1.41 \\
Edited 
& 1.85 & 70.5 & 0.304 & 8.47 & 1.42
& 1.83 & 71.2 & 0.299 & 9.76 & 1.40 \\
AI-Generated 
& 1.78 & 69.4 & 0.305 & 11.69 & 1.41
& 1.61 & 81.4 & 0.362 & 11.30 & 1.26 \\
\hline
\end{tabular}
\end{table*}

Community Notes begin in a ``Needs More Ratings'' (NMR) state and transition to either ``Helpful'' or ``Not Helpful'' once sufficient cross-partisan agreement is reached, indicating community consensus.
Only notes that exit NMR are displayed under the flagged post. 
Using the `Note Status History' file, we compute consensus-related metrics at both the tweet and note level.
At the note level, we compute the (1) \textbf{Notes / Tweets}: the average number of notes per tweet, and (2) \textbf{Helpful (\%)}: the share of notes rated Helpful.
At the tweet level, we measure 
(3) \textbf{Consensus Probability}: the fraction of tweets receiving at least one note transition from NMR to consensus, 
(4) \textbf{First Note}: the median time (in hours) between tweet creation and its first associated note,  
and 
(5) \textbf{Notes to Consensus}: the average number of notes required before the first consensus note appears.

Table \ref{tab:consensus} summarizes the results. 
Across both modalities, AI-generated content receives fewer notes per tweet and takes longer to receive its first note (11.3h vs 9.1h in video).
However, once annotated, AI-related posts are more likely to reach consensus. 
In the video subset, AI-generated tweets exhibit a consensus probability of 0.362 compared to 0.295 and 0.299 for other categories, and require fewer notes on average before consensus (1.26 vs 1.40). 
Moreover, AI-generated content shows the highest Helpful share (81.4\% in video; 69.4\% in image).
These patterns indicate a distinct annotation dynamic for AI-generated visual content: it attracts fewer and slower annotations, yet once examined, raters converge more quickly and with higher agreement. 
In contrast, miscaptioned content receives more immediate attention but exhibits lower consensus rates and a smaller share of helpful outcomes.

One explanation is that current generative outputs 
contain identifiable structural artifacts that facilitate objective verification once attention is drawn.
Furthermore, the frequent citation of external verification tools, a phenomenon which we analyze in Section~\ref{sec:ai_signals}, provides a standardized evidence base that accelerates rater agreement.
Combined with the engagement results in Section \ref{sec:engagement}, this suggests that while synthetic misinformation exhibits higher virality through passive engagement, it undergoes swift collective correction once subjected to crowdsourced scrutiny.

\section{AI References in Community Notes}
\label{sec:ai_signals}

To quantify how AI is used and referenced in moderation discussions, 
we examine general references to AI generation (e.g., `AI generated', `created by artificial intelligence'), which attribute content creation to AI systems, 
and mentions of specific AI models or tools (e.g., ChatGPT, Grok, Gemini, Midjourney, Sora).
See Appendix 12
for further methodological details.

AI-related references are far more prevalent in Community Notes than in user posts; appearing in $\approx11\%$ of notes compared to less than 1\% of posts.
As shown in Table~\ref{tab:ai_mentions}, general references to AI concentrate strongly in notes addressing AI-generated misinformation. 
In the image subset, 60.6\% of notes attached to AI-generated content contain at least one AI-related reference (either specific AI model or general references to AI) compared to 2--3\% for miscaptioned and edited content. 
Similar patterns appear in the video subset, where over half of AI-related notes reference AI explicitly. 
However, a substantial fraction of notes addressing synthetic media, $\approx$40\% in the image subset, do not contain explicit AI-generation phrases. 
This suggests that identifying synthetic media often relies on contextual reasoning or visual forensics beyond simple keyword cues, which supports the adopted hybrid labeling approach described in Section~\ref{sec:dataset}.

\begin{table}[t]
\centering
\caption{AI-related mentions in Community Notes. Values represent the percentage of notes containing \emph{general} references to AI or to specific a \emph{model}.}
\label{tab:ai_mentions}

\resizebox{\columnwidth}{!}{
\begin{tabular}{l|ccc|ccc}
\hline
 & \multicolumn{3}{c|}{Image} & \multicolumn{3}{c}{Video} \\
Type & Model & General & Any & Model & General & Any \\
\hline
Miscaptioned & 0.94 & 1.58 & 2.48 & 0.98 & 1.72 & 2.68 \\
Edited & 1.35 & 1.47 & 2.80 & 0.92 & 1.63 & 2.53 \\
AI-generated & 9.47 & 56.75 & 60.58 & 12.25 & 47.07 & 53.04 \\
\hline
\end{tabular}
}

\end{table}

Explicit references to specific AI systems are less frequent than general references to AI but still present across the dataset. 
The most frequently mentioned systems are general-purpose assistants such as Grok (46.8\% in image set; 39.4\% in video set), ChatGPT (16\% in image set; 12.1\% in video set) and Gemini (14.5\% in image set; 4.6\% in video set), which appear across both modalities. 
Modality-specific tools are also referenced: in the image subset, mentions often include Midjourney (11.6\%), while in the video subset references are dominated by Sora (37.2\%) and Veo (4.1\%). 
Notably, a substantial fraction of model mentions appear inside URLs rather than descriptive text. 
For example, notes often include links to Grok-generated replies embedded on X or shared conversations with systems such as Grok or ChatGPT.
This suggests that community contributors are increasingly citing AI-assisted analyses as supporting evidence to justify the synthetic nature of a post.

\section{Evaluation of Detection Systems}
\label{sec:evaluation}

As AI systems are increasingly cited within the explanatory context of Community Notes, we examine how reliably current automated methods distinguish AI-generated from authentic images in the wild.

%------------------------------------------------------------------------
\begin{figure*}[t]
    \centering
    \includegraphics[width=\textwidth]{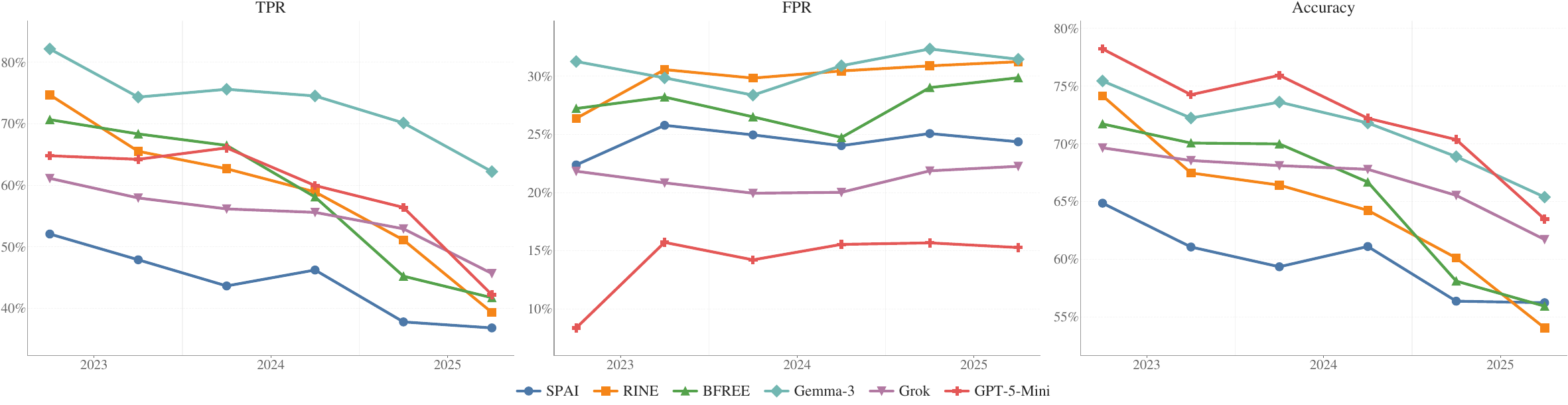}
    \caption{Temporal performance of AI-image detection systems from 2023 to 2025. Panels show TPR, FPR, and Accuracy across six-month evaluation periods. 
    TPR declines consistently across models, while FPR remains relatively stable, suggesting that performance degradation is driven by reduced sensitivity to newer AI-generated images.
    }
    \label{fig:benchmark_eval_overtime}
\end{figure*}

\subsection{Experimental Setup}

From \datasetname, we select all images assigned to the \emph{AI-generated class} from 2023 to 2025, yielding 10,866 images.
Earlier cases (2020--2022) and a small number of entries from early 2026 are excluded due to their very limited count.
To construct a balanced evaluation set, we sample an equal number of images from the \emph{Miscaptioned class}--real images presented in misleading contexts--to serve as the authentic ground truth. 
To control for temporal shifts in image quality and model capabilities, sampling is matched by year to the AI-generated set (1,524 images for 2023; 3,822 for 2024; 5,520 for 2025). 
The resulting test set contains approximately 21.7K images, balanced between AI-generated and authentic images, and is used exclusively to evaluate off-the-shelf models without further training.

We evaluate three Synthetic Image Detectors (SIDs), SPAI, RINE, and BFree, and three VLMs, Gemma~3~27B, Grok~4.1~Fast, and GPT-5-mini. 
We evaluate all VLMs in a zero-shot setting and employ SIDs using their original pretrained weights without additional fine-tuning.
For SPAI, we use a probability threshold of 0.5. 
For RINE, predictions at or above \texttt{MODERATE EVIDENCE} are classified as AI-generated. 
BFree outputs a single logit score per image; positive values are classified as AI-generated and negative values as non-AI.
For VLMs, we prompt the model to return a single label (AI or real) based on the image alone.
See Appendix 13 for details and prompts. 

We treat the AI-generated class as the positive class in all evaluations and report True Positive Rate (TPR), False Positive Rate (FPR), Precision, F1-score, and Accuracy.

\subsection{Overall Results}

\begin{table}[t]
\centering
\small
\setlength{\tabcolsep}{1.2pt}
\caption{Overall benchmark performance on the image dataset. 
\textbf{Bold} denotes best values and \underline{underlined} the second-best.
}
\label{tab:benchmark_overall}
\begin{tabular}{lccccc}
\hline
Model & TPR $\uparrow$ & FPR $\downarrow$ & Precision $\uparrow$ & F1 $\uparrow$ & Accuracy $\uparrow$ \\
\hline
SPAI           & 41.48 & 24.66 & 62.90 & 49.99 & 58.34 \\
RINE           & 52.75 & 30.54 & 63.75 & 57.73 & 61.03 \\
BFree          & 52.72 & 27.81 & 65.65 & 58.47 & 62.42 \\
Gemma~3~27B    & \textbf{70.21} & 31.00 & 69.51 & \textbf{69.86} & \underline{69.60} \\
Grok~4.1~Fast  & 52.28 & \underline{21.16} & \underline{71.35} & 60.35 & 65.51 \\
GPT-5-mini     & \underline{55.00} & \textbf{15.06} & \textbf{78.64} & \underline{64.73} & \textbf{69.91} \\
\hline
\end{tabular}
\end{table}

Table~\ref{tab:benchmark_overall} summarizes overall detection performance across all evaluated models.
VLMs consistently outperform the specialized SIDs across all reported metrics. 
GPT-5-mini achieves the highest overall accuracy (69.91\%), while BFree is the strongest among the specialized detectors (accuracy 62.42\%, F1 58.47\%). 
This indicates that state-of-the-art general-purpose VLMs can match or exceed the performance of dedicated SIDs in an ``in the wild'' setting.
The primary differences between models appear in the trade-off between TPR and FPR. 
Gemma~3 achieves the highest TPR (70.21\%) and F1-score (69.86\%), 
at the cost of a high FPR (31\%). 
In contrast, GPT-5-mini exhibits the opposite behavior: it achieves the lowest FPR (15.06\%) and the highest precision (78.64\%), while maintaining moderate TPR (55\%). Grok~4.1~Fast occupies a middle ground, with balanced TPR and FPR, while SIDs generally exhibit lower TPR with only moderate reductions in FPR.

\subsection{Temporal Degradation}

To examine how detection performance evolves over time, we evaluate each model across the 2023--2025 period using six-month bins. 
As Figure~\ref{fig:benchmark_eval_overtime} shows, 
across all models, TPR declines substantially over time, with drops ranging from roughly 16\% to 36\% between early 2023 and late 2025. 
The strongest degradation is observed for the specialized SIDs: RINE declines from 74.69\% in early 2023 to 39.34\% in late 2025, while BFree decreases from 70.66\% to 41.73\% over the same period. 
VLMs also exhibit declining TPR, though to a slightly lesser extent. 
For example, Gemma~3 drops from 82.15\% to 62.22\%, 
and Grok declines from 61.12\% to 45.67\%.
These TPR reductions also translate 
in decrease in overall accuracy across the same period. 
For instance, RINE drops from 74.16\% in early 2023 to 54.04\% in late 2025, while Gemma~3 drops from 75.44\% to 65.38\%.
In contrast, FPR remains relatively stable over time, fluctuating within a narrow range. This indicates that the performance degradation is primarily driven by reduced sensitivity to AI-generated images rather than by increased misclassification of authentic images.

This performance decline reflects a distribution shift caused by the rapid evolution of generative models. 
SIDs trained on earlier generations of synthetic data rely on visual cues that become less reliable as AI-generated images become more realistic and stylistically diverse, while VLMs, despite broader pretraining and stronger reasoning, are constrained by their training data cutoffs.
Additionally, given the training of VLMs on large-scale web data, it is possible that some evaluation images—or visually similar instances—were present in their training corpora, which may partially influence performance.
These results highlight that relying solely on static AI-image detectors is insufficient in rapidly evolving misinformation environments.

\begin{figure*}[t]
\centering
\includegraphics[width=\textwidth]{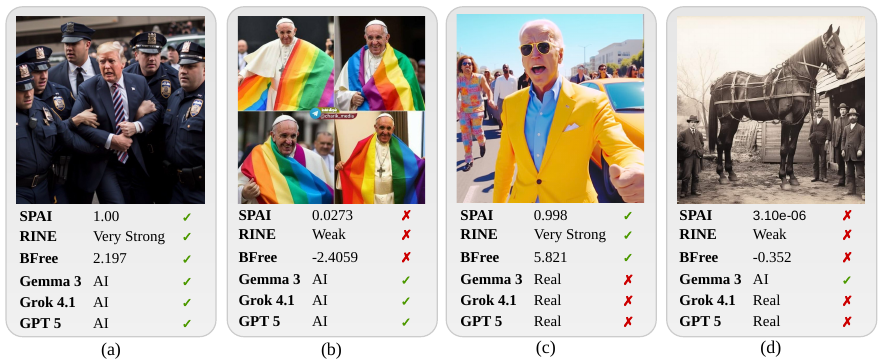}
\caption{Examples of AI-generated images with corresponding predictions from SIDs and VLMs.}
\label{fig:ai_examples}
\end{figure*}

\subsection{Qualitative Analysis}
Figure \ref{fig:ai_examples} shows four examples of AI-generated images and corresponding predictions by SIDs and VLMs.
In (a), all models correctly identify the image as synthetic, which may indicate the presence of noticeable artifacts alongside semantic inconsistencies that VLMs can potentially capture. 
Notably, the image is relatively older, posted in 2023, and 
it is possible that some models have been trained on it or similar images. In (b), all three SIDs fail while VLMs make the correct prediction. 
The image is relatively clean and lacks strong low-level artifacts, leading to weak SID signals, while its unusual or less plausible semantic content --namely, the pope with a rainbow flag-- enables VLMs to classify it as AI-generated. 
The opposite pattern appears in (c), where SIDs correctly detect the image as synthetic, possibly due to subtle artifact patterns, whereas VLMs incorrectly predict it as real; if the supposed identity of the depicted figure (former President Biden) is not recognized, the scene remains semantically plausible and does not raise obvious concerns. 
Finally, (d) presents a case where most models fail, potentially suggesting that the image resembles a degraded or historical photograph, thereby obscuring artifact-based cues.
Taken together, these examples suggest that SIDs and VLMs may rely on different and potentially complementary signals—low-level artifacts versus high-level semantics—which may account for their differing predictions across cases.

\section{Conclusion}
\label{sec:conclusion}

In this study, we present \datasetname, a large-scale dataset of multimodal misinformation, miscaptioned, edited, and AI-generated images and videos, collected from X's Community Notes. 
We leverage this data to conduct a longitudinal analysis of how misinformation evolves in terms of virality, engagement, consensus dynamics, and detectability.

Our results show that AI-generated visual content is undergoing a rise in volume as generative models evolve. 
While it achieves disproportionate virality, this spread is driven primarily by passive engagement (e.g., favorites) rather than the active discource typical of miscaptioned media. 
Furthermore, despite slower initial reporting, AI-generated visuals reach community consensus more quickly than other categories. 
This suggests that synthetic content currently possesses recognizable artifacts or standardized cues that facilitate collective verification--a process increasingly aided by the integration of AI-detection tools within the crowd-sourced annotation process.

To explore the reliability of specialized synthetic image detectors and VLMs, we create an evaluation benchmark of authentic vs. AI-generated images. 
Our evaluation uncovers a significant decline in True Positive Rate over time. 
This highlights the vulnerability of static detection systems, which struggle to generalize as generative models become more realistic and closer to authentic imagery.

The dissemination of AI-generated content and misinformation represents a dynamic challenge within a rapidly shifting digital environment. 
Due to the quick evolution of generative capabilities, this landscape necessitates a human-in-the-loop approach that combines community monitoring with automated fact-checking. 
Such efforts require constant improvement through the iterative re-training of detection models to keep pace with advancements in generative AI.

While this study offers a snapshot of the current state of the field, our proposed pipeline is designed for long-term analysis. 
It can be used to continuously monitor the evolution of synthetic media as new Community Notes data becomes available, providing researchers and platforms with real-time insights into multimodal misinformation.

\section*{Acknowledgments}
\label{sec:acknowledgements}

This work received funding by the Horizon Europe projects AI-CODE (grant agreement no. 101135437) and AI4Trust (101070190).

{
    \small
    \bibliographystyle{ieeenat_fullname}
    \bibliography{main}
}

\end{document}